\documentclass[fleqn,twoside]{article}
\usepackage{espcrc2}

\usepackage{graphicx}

\def\bge{\begin{equation}}
\def\ene{\end{equation}}
\def\bgea{\begin{eqnarray}}
\def\enea{\end{eqnarray}}
\def\nn{\nonumber}
\def\rxt{${\rm R\chi T}\,$}

\newcommand{\AmS}{{\protect\the\textfont2
  A\kern-.1667em\lower.5ex\hbox{M}\kern-.125emS}}

\title{Radiative decays with scalar mesons $a_0(980)$ and $f_0(980)$ in Resonance Chiral Theory}

\author{
  S.~Ivashyn\address[KIPT]{
      Institute for Theoretical Physics, NSC ``Kharkov Institute of Physics and Technology'',
      \\
      1, Akademicheskaya str., Kharkov 61108, Ukraine
    }\thanks{Partly supported by
            NASU grant for young researchers (contract~8.63/2008) and
            Joint NASU-RFFR scientific project N~38/50-2008.
            }
        and A.~Korchin\addressmark[KIPT]\thanks{
    Supported by the INTAS grant 05-1000008-8328.
        }
  }

\begin{document}

\begin{abstract}
The $\phi(1020)$ radiative decays form a good playground for study
of the low-lying $a_0(980)$ and $f_0(980)$ scalar mesons. The
complicated interplay of kaon loop mechanism, isoscalar meson
mixing and momentum dependence of the effective couplings
manifests itself in the invariant mass distributions in the
$\phi(1020) \to \pi^0 \pi^0 \gamma $ and $\pi \eta \gamma $
spectra. These distributions are fitted in framework of Resonance
Chiral Theory.
\end{abstract}

\maketitle


\section{Introduction}

The $e^+ e^-$ experiments in
Novosibirsk~\cite{Achasov:2000ym,Achasov:2000ku}
and Frascati~\cite{Aloisio:2002bt,Bini:2002uk}
allow one to study the $\phi(1020) \to \pi \pi \gamma $ (and $\pi
\eta \gamma $) radiative decays. The invariant mass distributions
of the pseudoscalar ($P$) pairs in these decays are of
considerable interest. The scalar mesons ($S$) $f_0(980)$ and
$a_0(980)$ are important intermediate resonances in the $\pi\pi$
and $\pi\eta$ channel correspondingly, and thereby they show up in
these spectra.

\begin{figure}
\label{fig:e+e-scalar-scheme}
  \includegraphics[height=.15\textwidth]{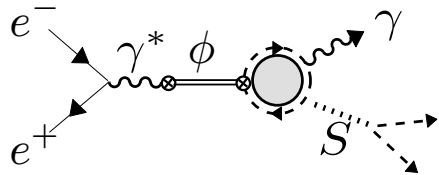}
  \caption{The $e^+e^-$ annihilation to $\pi\pi\gamma $ (and $\pi\eta\gamma$) via the $\phi$ vector
  resonance.
  $S$ denotes the intermediate scalar resonance. Blob shows schematically the kaon loop mechanism.}
\end{figure}

It is believed that the kaon loop (KL) coupling of $S$ to the
vector meson $\phi$ is very important (see, e.g. the data
analysis~\cite{Achasov:2000ym,Achasov:2005hm,Bugg:2006sr}).
Fig.\ref{fig:e+e-scalar-scheme} shows schematically the processes
with scalar mesons. Although many authors relate the dominance of
the KL mechanism to a large $K\bar{K}$ component in the $a_0(980)$
and $f_0(980)$ mesons and to the proximity of the $K\bar{K}$
threshold to the scalar meson mass, in fact the KL mechanism is a
feature of the chiral dynamics and reflects the important role of
the pseudoscalar mesons in low- and intermediate-energy
interactions. It is shown in
Refs.~\cite{Ivashyn:2007yy,KorchinIvashynLISBON2008} that the kaon
loops in the $\phi \to S \gamma$ transitions naturally arise in
the leading order in Resonance Chiral Theory
(\rxt)~\cite{EckerNP321}, irrespectively of the threshold and mass
position, and the internal structure of the scalars. The loop
contribution is convergent, gauge invariant and universal: one can
use either the analytical expression for $I(a,b)$~\cite{Close93},
or calculate it numerically.

The $\phi$ decays are rather suitable for study of the chiral
dynamics, and the framework is outlined in
Section~\ref{sec:Framework}. We should note that \rxt leads to an
important feature -- momentum dependence of the scalar meson
effective couplings. Employing the \rxt Lagrangian
of~\cite{EckerNP321} we obtained~\cite{Ivashyn:2007yy} a complete
set of $\mathcal{O}(p^4)$ contributions to various radiative
decays with the scalar mesons. Later on this model was
applied~\cite{KorchinIvashynLISBON2008} to the study of
$\pi^0\pi^0$ and $\pi^0\eta$ invariant mass distributions. In
Ref.~\cite{Ivashyn:2007yy,KorchinIvashynLISBON2008} we compared
the model results with estimates of other related
approaches~\cite{Black:2002ek,Nagahiro:2008mn}.

The free parameters of the model are $c_d$, $c_m$, $\tilde{c}_d$
and $\tilde{c}_m$, which are the coupling constants in the
Lagrangian~\cite{EckerNP321}. It is possible to fix their values
from the limit of large number of quark colors ($N_c \to \infty$),
 and by imposing the constraints suggested in
Ref.~\cite{Jamin:2001zq}. In the present paper we use the
corresponding values in the description of invariant mass
distributions in $\phi(1020) \to \pi^0 \pi^0 \gamma $ and $\pi
\eta \gamma $ decays.

The application of the model is presented in Section~\ref{sec:application}
and future prospects are outlined in Section~\ref{sec:prospects}.



\section{Scalar meson vertices and propagators in \rxt}
\label{sec:Framework}

\begin{figure}
  \includegraphics[width=.47\textwidth]{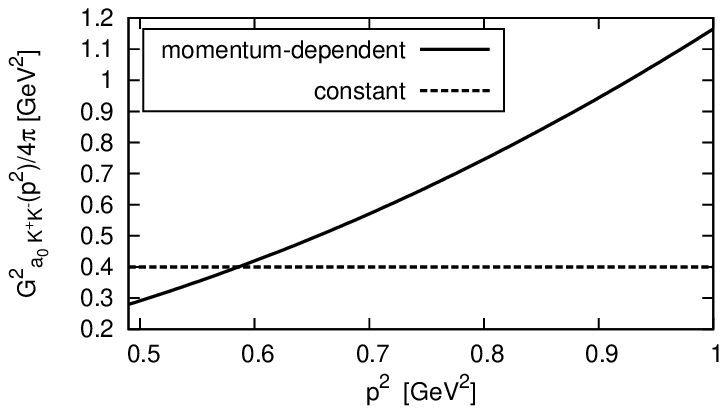}
  \\
  \includegraphics[width=.47\textwidth]{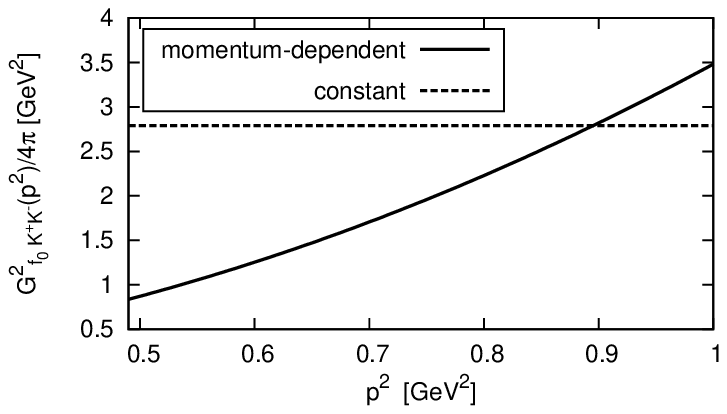}
  \vspace{-10pt}
  \label{fig:g2}
  \caption{
  Effective couplings $G_{a_0 K^+K^-}^2(p^2)/4\pi$ (upper panel)
  and $G_{f_0 K^+K^-}^2(p^2)/4\pi$ (lower panel).
  The values $c_m = c_d = f_{\pi}/2\;\approx 46.2$~MeV~\cite{Jamin:2001zq} are chosen.
  Mixing angle $\theta = 27.38$ is taken from our fit with Flatt\'e-like propagators.
  The horizontal lines indicate the constant values,
  see~\cite{Bini:2002uk}.
}
\end{figure}

In the \rxt a scalar meson $S(p)$ decays into the pseudoscalar
meson pair $P_1(p_1)\, P_2(p_2)$ through the momentum-independent
vertex ${i} {{c}_{SPP}}/{f_\pi^2}$ and the momentum-dependent one
${i} {\hat{c}_{SPP}(p_1\cdot p_2)}/{f_\pi^2}$ (see
Table~\ref{table:generalscalarcouplings}), where $f_\pi =
92.4~$MeV. These vertices vanish in the chiral limit, as soon as
${c}_{SPP}$ is proportional to the pseudoscalar meson mass
squared. The $f_0\to \pi\pi$ tree-level decay width is then
 \bgea
\label{width:fpp}
\Gamma_{f_0\to \pi\pi}(p^2) &=& \tilde{\Gamma}_{f_0\to \pi\pi}(p^2)\; {\rm \Theta}(p^2 - 4 m_\pi^2),
\nn
 \enea
where ${\rm \Theta}(x)$ is the Heaviside step function. For the
further convenience we employ the analytic function of $p^2$
 \bgea
\label{width:fpp:flatte} \tilde{\Gamma}_{f_0\to\pi\pi}(p^2) \!\!
&\!\!=\!\!&\!\!\frac{3}{2}\, \frac{1}{2 p^2} \sqrt{\frac{p^2}{4} -
m_\pi^2} \frac{G_{f_0\pi\pi}^2(p^2)}{4\pi},
 \enea
which is defined above the threshold and below the threshold with
$\sqrt{{p^2}/{4} - m_\pi^2} = i \sqrt{|{p^2}/{4} - m_\pi^2|}$. The
factor $3/2$ accounts for both $\pi^+\pi^-$ and $\pi^0\pi^0$ final
states and $G_{f_0\pi\pi}$ is a coupling of the $f_0$ to
$\pi^+\pi^-$.
 The analogous formulae hold for other decays. We
suppose that the scalar iso-singlet $f_0(980) = (S^{sing}\, \cos
\theta - S_8^{oct}\, \sin \theta)$
 is a mixture of the singlet $S^{sing}$ and the eights component of octet $S_8^{oct}$
with the angle~$\theta$.

It is important to note that the scalar meson couplings
${G_{SPP}^2(p^2)}/{4 \pi}$ are the essentially momentum-dependent
functions, see Table~\ref{table:generalscalarcouplings} for
explicit expressions. As an example, in Fig.~\ref{fig:g2} we
compare the functions ${G_{a_0 K^+ K^-}^2(p^2)}/{4 \pi}$ and
${G_{f_0 K^+ K^-}^2(p^2)}/{4 \pi}$ to the corresponding constant
values of these couplings, which are often discussed.

The total widths $\Gamma_{S,\; {tot}}(p^2)$ of the scalar mesons
are
\bgea
\label{width:total:f0}
\Gamma_{f_0,\, tot}(p^2) &=& \Gamma_{f_0\to \pi\pi}(p^2) + \Gamma_{f_0\to K\bar{K}}(p^2),
\\
\label{width:total:a0}
\Gamma_{a_0,\, tot}(p^2) &=& \Gamma_{a_0\to \pi^0\eta}(p^2) + \Gamma_{a_0\to K\bar{K}}(p^2).
\enea

\begin{table}
\caption{\rxt effective couplings (see Table~9 in Ref.~\cite{Ivashyn:2007yy}).}
\label{table:generalscalarcouplings}
\begin{tabular}{rcl}
\hline\noalign{\smallskip}
    $c_{f\pi\pi}$ &=& $- m_\pi^2 (4 \tilde{c}_m \cos \theta - 2\sqrt{2/3}\, c_m \sin \theta)$, \\
    $c_{fKK}$ &=& $- m_K^2(4 \,\tilde{c}_m \cos \theta + \sqrt{2/3}\, c_m \sin \theta)$ .\\
\noalign{\smallskip}\hline\noalign{\smallskip}
    $\hat{c}_{f\pi\pi}$ &=& $4 \,\tilde{c}_d \cos \theta - 2\sqrt{2/3}\, c_d \sin \theta$, \\
    $\hat{c}_{fKK}$ &=& $4\, \tilde{c}_d \cos \theta + \sqrt{2/3}\, c_d \sin \theta$.
\\
\noalign{\smallskip}\hline\noalign{\smallskip}
    $c_{aKK}$            &=& $- \sqrt{2}\, c_m m_K^2$,\\
    $c_{a\pi\eta}$       &=& $-2 \mathcal{Z}\sqrt{2/3} \, c_m m_\pi^2$  .\\
\noalign{\smallskip}\hline\noalign{\smallskip}
    $\hat{c}_{aKK}$      &=& $\sqrt{2} c_d$,\\
    $\hat{c}_{a\pi\eta}$ &=& $2 \mathcal{Z} \sqrt{2/3} \, c_d$ .

\\
\noalign{\smallskip}\hline\noalign{\smallskip}
    $G_{f_0 KK}$ & $\equiv$ & $ \frac{1}{f_\pi^2} \left( \hat{c}_{f_0KK} (m_K^2 - {p^2}/{2}) + c_{f_0KK} \right)$, \\
    $G_{f_0 \pi \pi}$ & $\equiv$ & $\frac{1}{f_\pi^2} \left( \hat{c}_{f_0\pi \pi} (m_\pi^2 - p^2/2) + c_{f_0\pi \pi} \right)$,\\
    $G_{a_0 KK}$ &$\equiv$ & $\frac{1}{f_\pi^2} \left( \hat{c}_{a_0KK} (m_K^2 - p^2/2) + c_{a_0KK} \right)$, \\
    $G_{a_0 \pi \eta}$ &$\equiv$ & $\frac{1}{f_\pi^2} \left(
     \hat{c}_{a\pi\eta} (m_\eta^2 + m_\pi^2 - p^2)/2 + c_{a\pi\eta} \right).$
     \\
\noalign{\smallskip}\hline
\end{tabular}
\\[2pt]
The factor $\mathcal{Z}=\frac{\cos \theta_0 - \sqrt{2} \sin \theta_8}{\cos( \theta_8 - \theta_0)}
    \approx 1.53,$
where $\theta_0$ and $\theta_8$ are the $\eta$ - $\eta^\prime$ mixing angles.
\end{table}

In Fig.~\ref{fig:e+e-scalar-scheme} the mechanism of the
scalar-meson production is depicted. The scalar meson propagates
and decays to a pair of pseudoscalar mesons. The finite resonance
width effects in the invariant mass distributions for $\pi^0\pi^0$ and
$\pi^0\eta$ in the $\phi$ radiative decays are known to be
important~\cite{Oller:2002na}.

For the scalar meson propagator one
may use
 \bgea \label{scalar-propagator}
D_{S}(p^2)&=& [p^2 - m_S^2 + i \sqrt{p^2} \Gamma_{S,\; {tot}}(p^2)]^{-1},
 \enea
where the scalar meson mass is denoted by $m_S$. A more advanced
form of the propagator including both real and imaginary parts of
the scalar-meson self energy $\Pi_S(p^2)$ was proposed
recently~\cite{Achasov:2004uq}.
 As an option, in the present work
we use the so-called Flatt\'e-like form~\cite{Flatte:1976xu} the
scalar-meson propagator
 \bgea
\label{scalar-propagator-Flatte} D_{S}(p^2)&=& [p^2 - m_S^2 + i
\sqrt{p^2} \tilde{\Gamma}_{S,\; {tot}}(p^2)]^{-1},
 \enea
where
 \bgea
\label{width:total:f0:flatte} \tilde{\Gamma}_{f_0,\, tot}(p^2) &=&
\tilde{\Gamma}_{f_0\to \pi\pi}(p^2) + \tilde{\Gamma}_{f_0\to
K\bar{K}}(p^2),
\\
\label{width:total:a0:flatte} \tilde{\Gamma}_{a_0,\, tot}(p^2) &=&
\tilde{\Gamma}_{a_0\to \pi^0\eta}(p^2) + \tilde{\Gamma}_{a_0\to
K\bar{K}}(p^2),
 \enea
which is supported by the phenomenological analysis,
see~\cite{Bugg:2006sr}. The propagator in the form of
eq.(\ref{scalar-propagator-Flatte}) accounts for a contribution of
the $K\bar{K}$ channel to the imaginary part of the self energy
above the $K\bar{K}$ threshold. It also approximately describes
the real part of the self energy below the threshold via the
analytic continuation.


\section{Application to the radiative decays}
\label{sec:application}

\begin{figure*}
  \label{fig:Results}
  \vspace{40pt}
  \includegraphics[width=.5\textwidth]{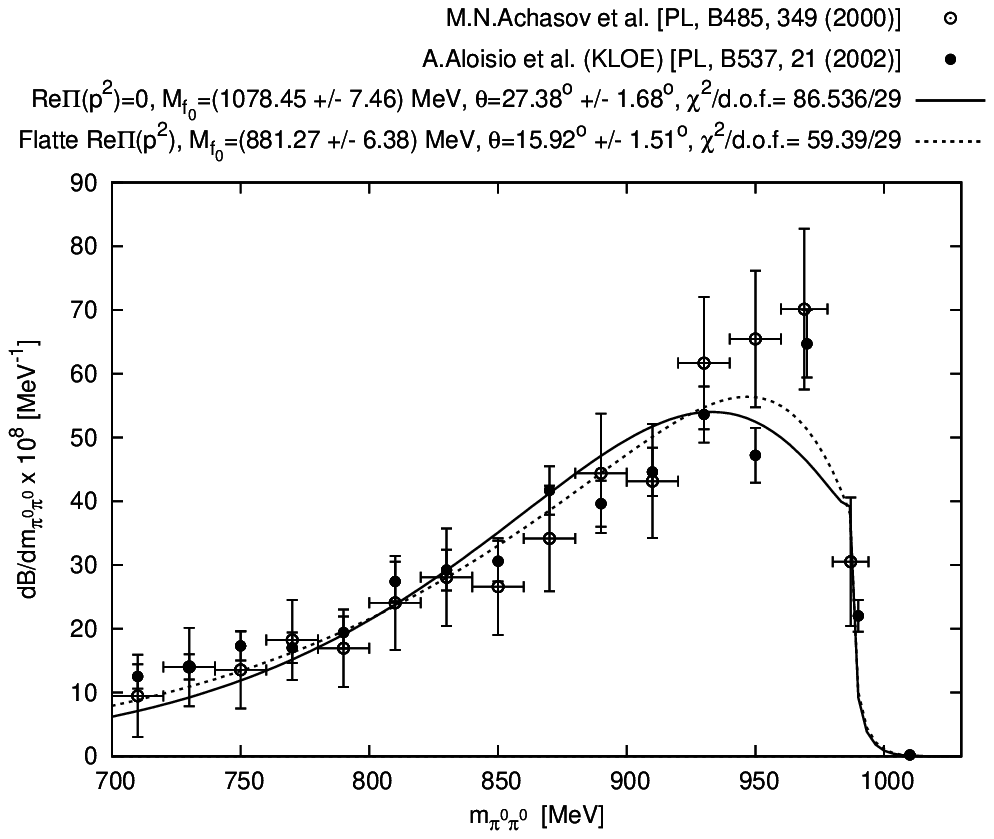}
  \includegraphics[width=.5\textwidth]{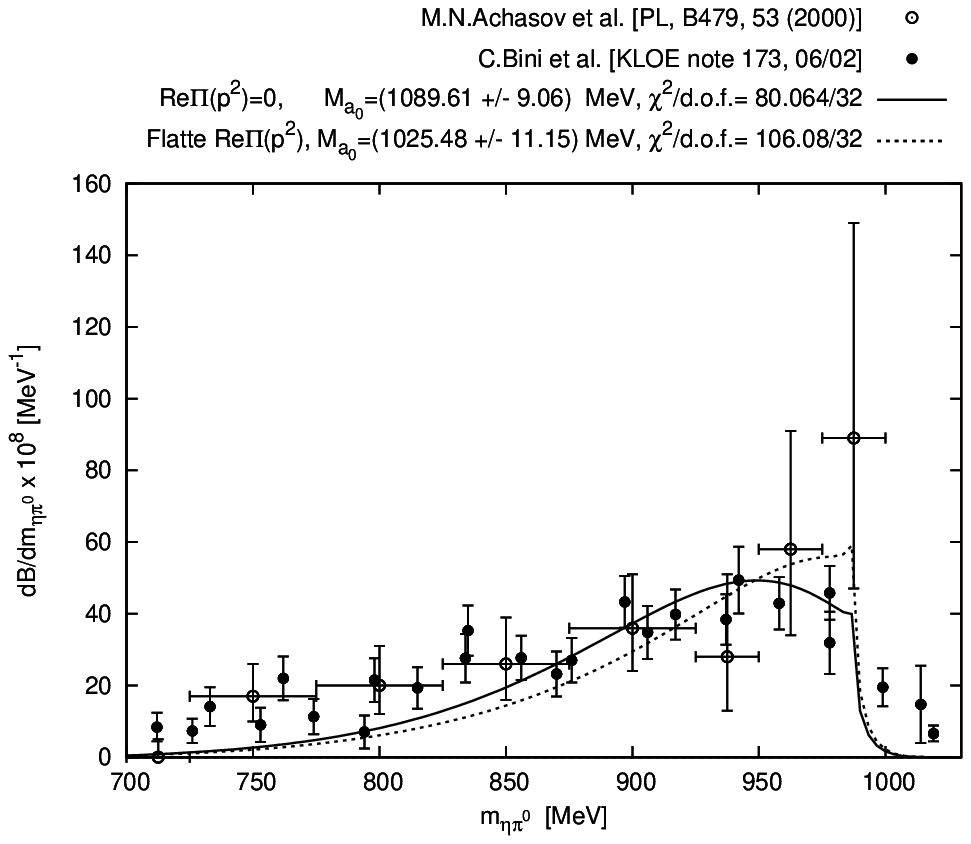}
  \caption{The invariant mass distributions in $\phi\to \pi^o\pi^o\gamma$ ({left panel})
and $\phi\to \pi^o\eta\gamma$ ({right panel}). Theoretical
calculation is performed with the values of parameters $c_m = c_d
= f_{\pi}/2\;\approx 46.2$~MeV~\cite{Jamin:2001zq}. The scalar
resonance masses and the mixing angle $\theta$ obtained from the fit are
shown in the legend. The two variants of the scalar meson
propagator are discussed in the text.
Data:~\cite{Achasov:2000ym,Aloisio:2002bt} (left)
and~\cite{Achasov:2000ku,Bini:2002uk} (right).}
\end{figure*}

A priori the scalar octet and singlet have independent couplings
$c_d, c_m$ and $\tilde{c}_d, \tilde{c}_m$ respectively. Numerical
values of these couplings, in principle, can be determined from the underlying
QCD. From the assumption $N_c \to\infty$ it was demonstrated~\cite{EckerNP321}
that the octet and
siglet (with ``tilde'') chiral couplings obey the constraints
 \bgea
 \label{eq:largeNC}
 \tilde{c}_m &=& \frac{c_m}{\sqrt{3}}, \;\;\;\;\;
 \tilde{c}_d = \frac{c_d}{\sqrt{3}}.
 \enea
In this limit the octet and singlet mesons become degenerate and
have equal masses.

We use the latter constraint although one may argue whether the
large-$N_c$ consideration is applicable to the scalar mesons,
especially in view of so-called Inverse Amplitude Method
results~\cite{Pelaez:2003dy}. The unusual large-$N_c$ behavior of
the scalar resonances was recently summarized
in~\cite{Jaffe:2007id}.

In Ref.~\cite{Jamin:2001zq} based on the short-distance
constraints on the flavor-changing $K \pi$, $K \eta$ and $K
\eta^\prime$ scalar form-factors it was shown that the values of
$c_m$ and $c_d$ couplings in the $\mathcal{O}(p^2)$ resonance
chiral Lagrangian satisfy the relation
 \bge \label{eq:jamin}
c_m = c_d = \frac{f_{\pi}}{2} \;\approx 46.2 \; {\rm MeV}.
 \ene
This relation, in particular, allows us to reduce the number of
adjustable parameters in the fit.

The expression for the invariant mass distribution in the KL model
is obtained in Ref.~\cite{Close93}, and its physical meaning is
explained in terms of the $SPP$ and $\phi PP$
couplings~\cite{Bugg:2006sr}. Taking into account the
momentum-dependent vertices we obtain
\bgea \label{eq:main} \!
 \frac{{\rm d}B}{{\rm
d}m_{\pi^0\pi^0}}\!\!\!&\!\!=\!\!&\!\!\! \frac{1}{2}
\frac{1}{\Gamma_{\phi,\,tot}} \frac{\alpha\,\sqrt{p^2}\,\sqrt{1 - 4
m_\pi^2/p^2}}{4\,\times\,48\pi^4 m_K^4}
\\
\nn && \!\!\!\!\!
\times\,\left|\frac{I(a,b)}{D_{f_0}(p^2)}\right|^2
\left(\frac{M_{\phi}^2 - p^2}{M_{\phi}}\right)^3
\\
\nn &&\!\!\!\!\! \times\, \frac{G_{f_0\pi\pi}^2(p^2)}{4\pi}
\frac{G_{f_0 KK}^2(p^2)}{4\pi}\left(\frac{\sqrt{2} G_V M_{\phi}
}{f_\pi^2}\right)^2
 \enea
for the $\pi^0 \pi^0$ invariant mass
($m_{\pi^0\pi^0} \equiv \sqrt{p^2}$)
distribution, here
$a=M_\phi^2/{m_K^2}$, $b=p^2/{m_K^2}$ and $\alpha\approx 1/137$.
The factor $1/2$ takes into account the identity of the neutral
pions.
 The formula for the $\pi^0 \eta$ case
is analogous to (\ref{eq:main}).

The $\phi(1020) \to \pi^0 \pi^0 \gamma $ (and $\pi \eta \gamma $)
spectra fitted by varying only the masses of scalar mesons and the
singlet-octet mixing angle $\theta$ are shown in
Fig.~\ref{fig:Results}.
We focus on the region near the $f_0(980)$ peak
and correspondingly use the data above $700\,$MeV.
The $\sigma=f_0(600)$ meson
is not included in the analysis.

 The figure also illustrates an effect from the
Flatt\'e-like modification of the scalar-meson propagator. The
shape of the $\pi^0 \pi^0$ and $\pi \eta$ distributions is well
reproduced. For the both forms of the
propagator~(\ref{scalar-propagator})
and~(\ref{scalar-propagator-Flatte}) the value of $\chi^2/d.o.f.$
is relatively small.

The mass of the $f_0(980)$ and the mixing angle $\theta$ turn out
to be strongly correlated in the fit. The masses of the scalar
resonances obtained from the fit with the propagator
(\ref{scalar-propagator}) are overestimated. We observe that the
account for the real part of the scalar-meson self energy, even in
a simple Flatt\'e-like approximation, results in a considerable
decrease of the $f_0$ mass in the fit. Therefore there is a strong
motivation for the careful treatment of the self-energy and the
effects related to a specific behavior of the scalar meson
spectral function. The importance of these aspects is discussed
also in Ref.~\cite{Oller:2002na}.

Usually the experimental results are interpreted in terms of the
constant values of the $SPP$ couplings $g_{SPP}^2/4\pi$ instead of
$G_{SPP}(p^2)$ functions. Constant $g_{SPP}$ couplings can be
justified in the phenomenological approaches, however this is not
supported by chiral models. An importance of the
derivative-coupling interactions was emphasized in
Ref.~\cite{Black:2002ek} some time ago, and recently discussed
in~\cite{Giacosa:2008st}. The replacement of $G_{SPP}(p^2)$
functions by the constant values may have a dramatic influence on
the invariant mass distributions. In fact, in these distributions
$p^2$ varies within the wide region -- from the $PP$ threshold up
to the $\phi$ mass squared -- and the scalar resonances may
``feel'' the variation of $G_{SPP}(p^2)$ along the spectrum. This
becomes more manifest in the case of a broad $\sigma$ meson, which
is sometimes included in the fit.

\section{Prospects and conclusion}
\label{sec:prospects}

We fit the $\phi(1020) \to \pi^0 \pi^0 \gamma $ and $\pi \eta
\gamma $ spectra by varying only the masses of scalar mesons and
the singlet-octet mixing angle $\theta$. The $\sigma=f_0(600)$
meson is not included yet. It is interesting that the strict
constraint $c_m = c_d = f_{\pi}/2\;\approx
46.2$~MeV~\cite{Jamin:2001zq} still leaves a possibility to fit
the data with a reasonably small $\chi^2/d.o.f$. At the same time
one realizes that the subtraction of a rather nontrivial
non-scalar resonance contributions may be very important for a
study of scalar resonance features.

Although our fit gives somewhat overestimated values for the mass
of $f_0(980)$ scalar resonance, there may be a way out. An
inclusion of the self-energy in the scalar meson propagator in an
advanced form may result in the mass lowering.

The dominant scalar resonance contribution of
Fig.\ref{fig:e+e-scalar-scheme} exhibits just a part of the $e^+
e^- \to \pi\pi\gamma$ and $e^+ e^- \to \pi\eta\gamma$ mechanisms.
The other contributions, which interfere with the scalar meson
contribution, are in general complicated. For example, in the case
of the neutral particles in the final state there are
$\phi\to\rho^0\pi^0\to\pi^0\pi^0\gamma$, \
$\phi\to\omega\eta\to\eta\pi^0\gamma$, \ $\phi
\to\rho^0\pi^0\to\eta\pi^0\gamma$ and other processes. Such
non-scalar resonance channels and background are important for the
precise analysis of the data (see, for example,
Ref.~\cite{Bugg:2006sr}). These contributions can also be included
in framework of \rxt with a relatively few number of free
parameters. The approach can be extended to a wide interval of the
$e^+ e^-$ center-of-mass energy, covering not only the $\phi$
resonance~\cite{Shekhovtsova}.

The model allows for an extension to the charged particles
production in $e^+ e^-$ annihilation. In addition to the $\phi$
decays, the framework is useful for a detailed study of other
radiative decays involving the light scalar mesons. Among the most
interesting ones there are the radiative decays into vector mesons
$f_0/a_0\to \gamma V$, $V=\rho,\omega$, currently under study in
J\"ulich~\cite{Buscher:2008ge}.

To summarize, the \rxt approach for the decays with scalar mesons
at order $p^4$ is outlined. Though it does not specify the
internal structure of scalar mesons, the important features of
these resonances are reproduced. The model allows to investigate a
complicated interplay of kaon loop mechanism, scalar-meson mixing
and momentum dependence of the effective couplings in the
invariant mass distributions.




\end{document}